\documentclass[oneside,12pt]{amsart}

\usepackage{times}
\usepackage{geometry}
\usepackage[utf8]{inputenc}
\usepackage{enumitem,amssymb}
\usepackage{ragged2e}
\usepackage{graphicx}
\usepackage{natbib,latexsym,url,enumitem,pdfpages,multicol}
\usepackage{color}
\usepackage{wrapfig}
\usepackage{caption}
\usepackage{setspace}
\newlist{thematic}{itemize}{8}
\setlist[thematic]{label=$\square$}
\usepackage{pifont}
\definecolor{todo}{RGB}{200,0,0}

\makeatletter
\let\jnl@style=\rm
\def\ref@jnl#1{{\jnl@style#1}}

\def\ref@jnl#1{{\jnl@style#1}}% 
\newcommand\aj{\ref@jnl{AJ}}%        % Astronomical Journal 
\newcommand\araa{\ref@jnl{ARA\&A}}%  % Annual Review of Astron and Astrophys 
\newcommand\apj{\ref@jnl{ApJ}}%    % Astrophysical Journal ++
\newcommand\apjl{\ref@jnl{ApJL}}     % Astrophysical Journal, Letters 
\newcommand\apjs{\ref@jnl{ApJS}}%    % Astrophysical Journal, Supplement 
\newcommand\ao{\ref@jnl{ApOpt}}%   % Applied Optics ++
\newcommand\apss{\ref@jnl{Ap\&SS}}%  % Astrophysics and Space Science 
\newcommand\aap{\ref@jnl{A\&A}}%     % Astronomy and Astrophysics 
\newcommand\aapr{\ref@jnl{A\&A~Rv}}%  % Astronomy and Astrophysics Reviews 
\newcommand\aaps{\ref@jnl{A\&AS}}%    % Astronomy and Astrophysics, Supplement 
\newcommand\azh{\ref@jnl{AZh}}%       % Astronomicheskii Zhurnal 
\newcommand\baas{\ref@jnl{BAAS}}%     % Bulletin of the AAS 
\newcommand\icarus{\ref@jnl{Icarus}}% % Icarus
\newcommand\jrasc{\ref@jnl{JRASC}}%   % Journal of the RAS of Canada 
\newcommand\memras{\ref@jnl{MmRAS}}%  % Memoirs of the RAS 
\newcommand\mnras{\ref@jnl{MNRAS}}%   % Monthly Notices of the RAS 
\newcommand\pra{\ref@jnl{PhRvA}}% % Physical Review A: General Physics ++
\newcommand\prb{\ref@jnl{PhRvB}}% % Physical Review B: Solid State ++
\newcommand\prc{\ref@jnl{PhRvC}}% % Physical Review C ++
\newcommand\prd{\ref@jnl{PhRvD}}% % Physical Review D ++
\newcommand\pre{\ref@jnl{PhRvE}}% % Physical Review E ++
\newcommand\prl{\ref@jnl{PhRvL}}% % Physical Review Letters 
\newcommand\pasp{\ref@jnl{PASP}}%     % Publications of the ASP 
\newcommand\pasj{\ref@jnl{PASJ}}%     % Publications of the ASJ 
\newcommand\qjras{\ref@jnl{QJRAS}}%   % Quarterly Journal of the RAS 
\newcommand\skytel{\ref@jnl{S\&T}}%   % Sky and Telescope 
\newcommand\solphys{\ref@jnl{SoPh}}% % Solar Physics 
\newcommand\sovast{\ref@jnl{Soviet~Ast.}}% % Soviet Astronomy 
\newcommand\ssr{\ref@jnl{SSRv}}% % Space Science Reviews 
\newcommand\zap{\ref@jnl{ZA}}%       % Zeitschrift fuer Astrophysik 
\newcommand\nat{\ref@jnl{Nature}}%  % Nature 
\newcommand\iaucirc{\ref@jnl{IAUC}}% % IAU Cirulars 
\newcommand\aplett{\ref@jnl{Astrophys.~Lett.}}%  % Astrophysics Letters 
\newcommand\apspr{\ref@jnl{Astrophys.~Space~Phys.~Res.}}% % Astrophysics Space Physics Research 
\newcommand\bain{\ref@jnl{BAN}}% % Bulletin Astronomical Institute of the Netherlands 
\newcommand\fcp{\ref@jnl{FCPh}}%   % Fundamental Cosmic Physics 
\newcommand\gca{\ref@jnl{GeoCoA}}% % Geochimica Cosmochimica Acta 
\newcommand\grl{\ref@jnl{Geophys.~Res.~Lett.}}%  % Geophysics Research Letters 
\newcommand\jcp{\ref@jnl{JChPh}}%     % Journal of Chemical Physics 
\newcommand\jgr{\ref@jnl{J.~Geophys.~Res.}}%     % Journal of Geophysics Research 
\newcommand\jqsrt{\ref@jnl{JQSRT}}%   % Journal of Quantitiative Spectroscopy and Radiative Trasfer 
\newcommand\memsai{\ref@jnl{MmSAI}}% % Mem. Societa Astronomica Italiana 
\newcommand\nphysa{\ref@jnl{NuPhA}}%     % Nuclear Physics A 
\newcommand\physrep{\ref@jnl{PhR}}%       % Physics Reports 
\newcommand\physscr{\ref@jnl{PhyS}}%        % Physica Scripta 
\newcommand\planss{\ref@jnl{Planet.~Space~Sci.}}%  % Planetary Space Science 
\newcommand\procspie{\ref@jnl{Proc.~SPIE}}%      % Proceedings of the SPIE 

\newcommand\actaa{\ref@jnl{AcA}}%  % Acta Astronomica
\newcommand\caa{\ref@jnl{ChA\&A}}%  % Chinese Astronomy and Astrophysics
\newcommand\cjaa{\ref@jnl{ChJA\&A}}%  % Chinese Journal of Astronomy and Astrophysics
\newcommand\jcap{\ref@jnl{JCAP}}%  % Journal of Cosmology and Astroparticle Physics
\newcommand\na{\ref@jnl{NewA}}%  % New Astronomy
\newcommand\nar{\ref@jnl{NewAR}}%  % New Astronomy Review
\newcommand\pasa{\ref@jnl{PASA}}%  % Publications of the Astron. Soc. of Australia
\newcommand\rmxaa{\ref@jnl{RMxAA}}%  % Revista Mexicana de Astronomia y Astrofisica

\newcommand\maps{\ref@jnl{M\&PS}}% Meteoritics and Planetary Science
\newcommand\aas{\ref@jnl{AAS Meeting Abstracts}}% American Astronomical Society Meeting Abstracts
\newcommand\dps{\ref@jnl{AAS/DPS Meeting Abstracts}}% American Astronomical Society/Division for Planetary Sciences Meeting Abstracts

\begin{document}
\noindent {\huge Astro2020 APC White Paper } \\[20pt]
\noindent {\LARGE FOBOS: A Next-Generation Spectroscopic Facility }\\

% FOBOS: A Next-Generation Spectroscopic Facility at the W. M. Keck Observatory
\noindent \textbf{Thematic Areas:} Project Paper
  
\noindent \textbf{Principal Author:}

\noindent Name:	Kevin Bundy \\
\noindent Institution:  University of California Observatories \\
\noindent Email:  kbundy@ucolick.org \\
\noindent Phone:  831-459-3539 \\
 
\noindent \textbf{Co-authors:} {\footnotesize K.~Westfall (UCO), N.~MacDonald (UCO), R.~Kupke
(UCO), M.~Savage (UCO),
C.~Poppett (UCB/SSL), A.~Alabi (UCSC), G.~Becker
(UCR), J.~Burchett (UCSC), P.~Capak (Caltech), A.~Coil (UCSD),
M.~Cooper (UCI), D.~Cowley (UCO), W.~Deich (UCO), D.~Dillon (UCO),
J.~Edelstein (LBNL), P.~Guhathakurta (UCSC), J.~Hennawi (UCSB), M.~Kassis (WMKO),
K.-G.~Lee (IPMU), D.~Masters (JPL), T.~Miller (UCB/SSL), J.~Newman
(Pitt), J.~O'Meara (WMKO), J.~X.~Prochaska (UCSC), M.~Rau (CMU), J.~Rhodes (JPL), R.~M.~Rich (UCLA),
C.~Rockosi (UCSC), A.~Romanowsky (SJSU/UCSC), C.~Schafer (CMU), D.~Schlegel (LBNL),
A.~Shapley (UCLA), B.~Siana (UCR), Y.-S.~Ting (IAS), D.~Weisz
(UCB), M.~White (UCB/LBNL), B.~Williams (UW), G.~Wilson (UCR),
M.~Wilson (LBNL), \& R.~Yan (UK)} \\

% K. Westfall, N. MacDonald, R. Kupke, M. Savage,
% C. Poppett, A. Alabi, G. Becker, J. Burchett, P. Capak, A. Coil,
% M. Cooper, D. Cowley, W. Deich, D. Dillon,
% J. Edelstein, P. Guhathakurta, J. Hennawi, M. Kassis,
% K.-G. Lee, D. Masters, T. Miller, J. Newman, J. O'Meara, J. X. Prochaska, M. Rau, J. Rhodes, R. M. Rich,
% C. Rockosi, A. Romanowsky, C. Schafer, D. Schlegel,
% A. Shapley, B. Siana, Y.-S. Ting, D. Weisz, M. White, B. Williams, G. Wilson,
% M. Wilson, R. Yan

\noindent \textbf{Abstract:} 

\noindent High-multiplex and deep spectroscopic follow-up of upcoming
panoramic deep-imaging surveys like LSST, Euclid, and WFIRST is a widely recognized and increasingly urgent necessity.
No current or planned facility at a U.S.~observatory meets the sensitivity, multiplex, and rapid-response time needed
to exploit these future datasets. FOBOS\footnote{\url{fobos.ucolick.org}}, the Fiber-Optic Broadband Optical Spectrograph, is a near-term fiber-based
facility that addresses these spectroscopic needs by optimizing depth over area and exploiting the aperture advantage
of the existing 10m Keck II Telescope. The result is an instrument with a uniquely blue-sensitive wavelength range
(0.31--1.0 $\mu$m) at $R \sim 3500$, high-multiplex (1800 fibers), and a factor 1.7 greater survey speed and
order-of-magnitude greater sampling density than Subaru's Prime Focus Spectrograph (PFS). In the era of panoramic deep
imaging, FOBOS will excel at building the deep, spectroscopic reference data sets needed to interpret vast imaging
data. At the same time, its flexible focal plane, including a mode with 25 deployable integral-field units (IFUs)
across a 20 arcmin diameter field, enables an expansive range of scientific investigations. Its key programmatic areas
include (1) nested stellar-parameter training sets that enable studies of the Milky Way and M31 halo sub-structure, as
well as local group dwarf galaxies, (2) a comprehensive picture of galaxy formation thanks to detailed mapping of the
baryonic environment at $z \sim 2$ and statistical linking of evolving populations to the present day, and (3) dramatic
enhancements in cosmological constraints via precise photometric redshifts and determined redshift distributions.  In
combination with Keck I instrumentation, FOBOS also provides instant access to medium-resolution spectroscopy for
transient sources with full coverage from the UV to the K-band.

\pagebreak

%% Executive Summary and Overview
%\input{Astro2020-Overview}
%%%%
% -- Overview Material
%%%%

\vspace{-0.5cm}
\section{Scientific Motivation}

\subsection{Community Need} The need for spectroscopic follow-up in
the LSST era was made clear in the National Research Council's 2015
report, ``Optimizing the U.S. Ground-Based Optical and Infrared
Astronomy System'' \citep{NAP21722}:
\noindent\begin{center}\mbox{\parbox{0.9\linewidth}{
The National Science Foundation should support the development of a
wide-field, highly multiplexed spectroscopic capability on a medium- or
large-aperture telescope in the Southern Hemisphere to enable a wide
variety of science, including follow-up spectroscopy of Large Synoptic
Survey Telescope targets. Examples of enabled science are studies of
cosmology, galaxy evolution, quasars, and the Milky Way.
}}\end{center}

Workshops organized by the National Optical Astronomy Observatory
(NOAO) in 2013 and 2016 reported specific spectroscopic needs for
LSST follow-up in all science areas. In particular, the 2016 report
notes that a critical resource in need of prompt development is to
``Develop or obtain access to a highly multiplexed, wide-field
optical multi-object spectroscopic capability on an 8m-class
telescope.''  More recently, the need for significant investment in spectroscopic facilities was highlighted in multiple Astro2020 Science White Papers, many of which we refer to when motivating our science cases below.

FOBOS takes a critical first step in addressing these needs using an
existing telescope to achieve a final cost $\approx$20 times less
than wide-area spectroscopic telescopes of the future, such as the
Mauna Kea Spectroscopic Explorer \citep[MSE,][]{mse2018} and SpecTel
(see Astro2020 APC White Paper). Compared to the Prime Focus
Spectrograph (PFS) on Japan's Subaru Telescope, FOBOS would be
1.7$\times$ faster, provide unique UV sensitivity (0.31--1 $\mu$m
compared to 0.38--1.25 $\mu$m with PFS), and offer higher-density,
more flexible target sampling over ``deep-drilling'' fields. Unlike
PFS, FOBOS would be operated on a U.S.\ telescope with dedicated
U.S.\ access and a commitment to supporting U.S.-led imaging
facilities.

\subsection{Key Science Goals}

With its high multiplex and Keck's large aperture, FOBOS will enable significant progress in multiple science areas by
providing much-needed large and deep spectroscopic samples.  These unprecedented data sets will be scientific gold
mines for the U.S.~community in their own right, but when combined with novel observations from forthcoming facilities,
transformational advances are possible.  These include 1) charting the assembly history of the Milky Way, M31, and
Local Group dwarf galaxies by combining deep FOBOS spectroscopy with wide spectroscopic campaigns (e.g., DESI
Bright-Time Survey and SDSS-V Milky Way Mapper), GAIA data, and panoramic imaging from LSST, Euclid, and WFIRST; 2)
mapping the baryonic ecosystem at $z \sim 2$--3 and linking it to evolving populations at lower redshifts by training
photometric diagnostics that transfer detailed spectroscopic knowledge to billion-plus galaxy samples provided by
future all sky surveys; 3) dramatically enhancing cosmological probes using panoramic deep imaging and
cross-correlation techniques with Stage-IV CMB observations thanks to precise calibration of photometric redshifts and
redshift distributions.

FOBOS addresses these goals by achieving high multiplex while optimizing for sensitivity over area.  Future imaging
data will routinely reach $i_{\rm AB} = 25$, yielding target densities of 42 arcmin$^{-2}$.  FOBOS achieves a
single-pass on-sky sampling density of 6 arcmin$^{-2}$ on average, and close packing would allow a maximum target density of
$\sim$30 arcmin$^{-2}$.  These capabilities allow FOBOS to collect large samples of very faint sources with highly
efficient observing strategies.

Meanwhile, innovations in astrostatistics and machine learning in particular promise powerful synergies between these
information-rich but volume-limited FOBOS samples and the vast cosmic volumes that will be surveyed by LSST, Euclid,
and WFIRST.  As we discuss below, these synergies underlie all three of FOBOS's broad science goals.  Progress in all areas can be carried out with observations that simultaneously integrate multiple programs and observing modes, including a priority on time-domain astronomy.  A rapid response capability for high-value transient sources is provided by FOBOS's fixed, oversized IFU.  With FOBOS on Keck II and existing Keck I instrumentation, Keck will enable instant follow-up spectroscopy at medium resolution with full wavelength coverage from the UV to the K-band.  

Given the community-wide value of these capabilities  and
following past examples at Keck \citep[e.g., DEEP2][]{newman13}, FOBOS will play a leading role in obtaining and
publicly delivering critical, enabling data sets that leverage major project investments by the U.S.~community.  These
data sets will include raw observations, fully-reduced, calibrated, and processed spectra, and high-level derived data
products, including redshifts, measurements of spectral features, and inferred physical information (e.g., stellar
parameters, galaxy star formation histories, environmental catalogs).  In addition to the 20\% of Keck observing time
already open to the public, additional FOBOS access may be possible through key programs allocated a specific fraction
of FOBOS fiber-hours and combined with other programs into multi-cycle observing campaigns.  The FOBOS team will engage in a broad campaign of community engagement in the next year to further define science goals and community access models.  We encourage interested parties to contact the first author.

%\input{Astro2020-LocalGroup}
%%%%
% -- Local Group Science Cases
%%%%

\subsection{Assembly History of the Local Group}
\label{sec:localgroup}

Studies of individual stars in the Milky Way (MW), Magellanic Clouds, Andromeda (M31), Triangulum galaxy (M33), and
numerous dwarf satellites provide an exquisitely detailed look at specific examples of galaxy assembly and evolution.
While Gaia provides on-sky motions and photometry for 1.7 billion stars in the MW, fewer than 10\%, 0.3\%, and 0.1\% of
stars will have a full complement of astrometrics and kinematics, basic stellar parameters, and chemical abundances,
respectively.  Moreover, Gaia distance errors increase quadratically with distance.  Spectroscopy with APOGEE, the
Milky Way Mapper, and WEAVE provide supporting wide-field data sets but accounting for fainter stars requires
FOBOS-like sensitivity \citep[see][]{dey19,sanderson19}.  By carefully exploiting the overlap in these data sets, FOBOS
can link high-resolution and robust stellar information from brighter targets to stars that can only be characterized
by photometry.  This would enable data-driven models capable of providing photometric estimates of stellar
parameters (temperature, surface gravity, metallicity, and alpha-element abundance) for {\it all} stars in the Gaia
dataset  \citep[see][]{2015ApJ...808...16N, 2018arXiv180401530T, 2018arXiv180803278T}.

Of particular interest is the ability of future imaging surveys to increase the census of stellar streams and other
substructure by a hundredfold.  The stars in these structures are faint, however, and easily confused with background
galaxies in ground-based photometry.  With spectrocopic reference samples from FOBOS, the goal is to photometrically
reconstruct the star-formation histories of disrupted satellites and compare them with dynamical models to constrain
assembly histories and enclosed mass constraints \citep[e.g.,][]{2017ApJ...836..234S}.

Performing a similar analysis on the M31 halo is highly desireable but more challenging because individual
main-sequence stars at the distance of M31 are too faint for 10m telescopes.  Thus spectroscopic training efforts must
focus on giant stars in the M31 halo and be calibrated with hydrodynamical simulations that account for M31's differing
formation history  \citep[e.g.][]{2005MNRAS.356.1071R,li19}.

%\input{Astro2020-Galaxies}
%%%%
% -- Galaxies Science Cases
%%%%

\subsection{A Comprehensive Picture of Galaxy Formation}

With both single-fiber and multiplexed IFU observations, FOBOS will produce rich and comprehensive data sets at faint source magnitudes.  Its blue sensitivity affords UV absorption studies down to $z \sim 1.5$, enabling detailed mapping of the baryonic environment at the peak formation epoch.  Samples at $z=1$--$2$ will not only characterize how this environment and its impact on galaxies evolves but will also provide large training sets that can be used to extract spectroscopic-like information from the billion-plus galaxy samples observed in all-sky surveys.  These data will be used in concert with large samples of spatially-resolved FOBOS observations (in IFU mode) to set the context for highly-detailed studies of targeted samples with James Webb Space Telescope and the U.S.~Extremely Large Telescopes.  Finally, FOBOS can tie evolutionary behavior seen at early times to the present day by observing faint sub-structure and dynamical tracers in nearby galaxies.

\begin{wrapfigure}{l}{0.6\textwidth}
\includegraphics[width=0.6\textwidth]{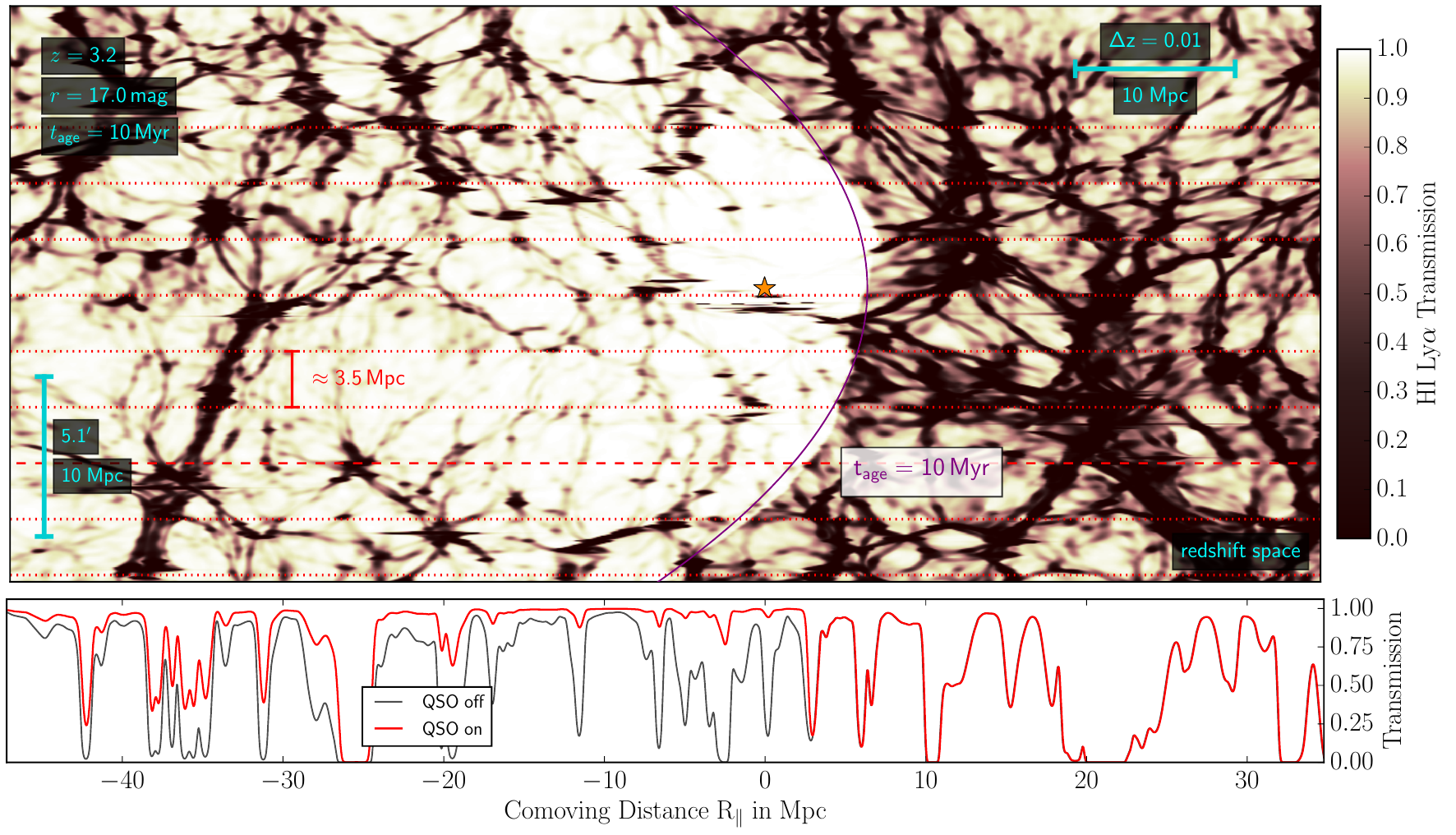}
\caption{{\it Top}: FOBOS will map quasar ``Light Echos'' as shown in this simulated
tomographic IGM map in the immediate environs of a quasar (gold star)
with several sightlines indicated
\citep[from][]{2018arXiv181005156S}. {\it Bottom}: The ionizing flux
within the echo's extent enhances transmission of Ly$\alpha$ photons
impinging on absorbers along the line-of-sight.}
\label{fig:LightEcho}
\end{wrapfigure}

\subsubsection{The $z$$\sim$2 galaxy ecosystem}
\label{sec:z2galaxies}

With publicly-accessible surveys like MOSDEF \citep{kriek15}, the Keck MOSFIRE instrument has provided powerful new
insights into early galaxies at the $z$$\sim$2 peak-formation epoch \citep[also see KBSS,][]{steidel14}. However, a complete
picture of the galaxy ``ecosystem'' at this key epoch must also consider the gas-filled environments. Using Ly$\alpha$
absorption in background galaxies, a tomographic map of the intergalactic medium (IGM) in regions surveyed by MOSDEF
and KBSS is a key first step. The promise of this approach, demonstrated at Keck by \citet{lee14}, motivates FOBOS's UV
sensitivity, target flexibility, and multiplex for tomographic mapping of large-scale structure, including
protoclusters \citep{lee16,kartaltepe19}, voids \citep{krolewski18}, and filaments \citep{horowitz19}.
\citet{2018arXiv181005156S} take IGM tomography in a new direction, demonstrating with simulated observations that
quasar ``light echos''
--- spatial signatures of the expanding ionization front of a newly
activated quasar --- can be detected and used to infer opening angles
and deconstruct the quasar's accretion history (see Fig
\ref{fig:LightEcho}). The required FOBOS spectra can simultaneously
constrain the CIV mass density (via $\lambda\lambda$1548,1550 \AA)
and patterns of CIV enrichment on both IGM and circumgalactic scales,
revealing the imprint of galaxy fueling and feedback processes
\citep[e.g.,][]{tumlinson17}.

\subsubsection{Role of environment at $z=1$--$2$ }

At later times, as IGM material becomes more confined to galaxies and their dark matter halos, these halos
increasingly merge to form larger structures.  FOBOS will be particularly important for mapping out environmental
effects at $z=1$--$2$ on galaxy evolution at the group scale ($\mathcal{M_\ast/M_\odot} \lesssim 10^{13}$) and
statistically linking galaxies to their host dark matter halos \citep{behroozi19}.  Tens
of thousands of satellites down to sub-L$^*$ luminosities will be mapped and characterized. Thanks to deep, wide-field
imaging surveys, like LSST, a 1M-object environmental survey at $z=1$--$2$ may then be possible using improved
photo-$z$s, strong priors on spectral types, and new machine-learning techniques to deliver {\it spectroscopic}
redshifts (with $\lesssim$300 km/s accuracy) at the lowest signal-to-noise possible (exposure times reduced by factors
of 4--5).

\subsubsection{Internal structure of galaxies at intermediate redshift}

MaNGA \citep{bundy15} and other large IFU surveys are defining the
$z=0$ benchmark for how internal structure is organized across the
galaxy population. To understand and test ideas for how this internal
structure emerged, we require spatially-resolved observations at $z =
1$--2, just after the peak formation epoch. Indeed, Keck has
helped pioneer such observations \citep[e.g.,][]{erb04, miller11,law09}.
With only one galaxy observed at a time, samples have so far been limited to a few hundred sources; however,
FOBOS will obtain resolved spectroscopy for thousands of galaxies
using its IFU-mode with a 25 deployed IFUs. Bright optical emission-line tracers
for this unprecedented sample of galaxies will reveal their gas-phase
and kinematic structure. Stacking restframe $\lambda \approx 4500$
spectra will enable radial stellar population analyses to constrain
how stellar components formed and assembled. Although initially
limited to coarse spatial scales, ground-layer adaptive optics (GLAO)
in combination with FOBOS would be transformative for this science. A
corrected FWHM of 0.2-0.3 arcsec would enable fine-sampling IFUs to
probe smaller galaxies and study sub-structure on 1--2 kpc scales.

\subsubsection{The dark and luminous content of nearby galaxies}

Environmental effects and evolutionary processes evident at higher redshift have consequences that can be tested in present-day galaxies.  Using globular cluster and planetary nebulae as tracers, FOBOS will
dramatically advance dynamical studies of nearby galaxies with
$\mathcal{M_\ast/M_\odot} \lesssim 10^{11}$, capturing the majority
of the $\sim$1000 GCs located within $\sim$50 kpc of typical hosts
\citep[see][]{2013ApJ...772...82H} and tightly constraining their
dark matter halos. FOBOS's multi-IFU mode will additionally provide
powerful insight on the origin of dwarf galaxies, both compact and
ultra-diffuse (UDGs), in the field and in nearby clusters like Coma
and Virgo.

%\input{Astro2020-Cosmology}
%%%%
% -- Cosmology Science
%%%%

\subsection{Dramatically Enhancing Cosmological Probes}

\subsubsection{Dark Energy Probes via Precision Cosmic Distances.}
\label{sec:cosmology}

Panoramic imaging surveys (e.g., LSST, Euclid, and WFIRST) are seeking to constrain the dark-energy
equation-of-state at $z \lesssim 1$ through measurements of angular correlations of galaxy positions, their
gravitational lensing shear, and the cross-correlation between the two.  These surveys rely on photometric redshifts
(``photo-$z$s''), whose uncertainties and potential biases are the major limitation and source of systematic error in
these efforts \citep{newman19,mandelbaum19}.  \citet{newman15} define a \emph{spectroscopic} survey for photo-$z$ training that would \emph{increase
the dark energy figure-of-merit in LSST by 40\%}.  The survey program is ideally matched to FOBOS.  It requires 10
independent fields, each 20 arcmin in diameter, with a sampling density of 6 arcmin$^{-2}$, and the ability to go very
deep ($i_{\rm AB} < 25.3$).  FOBOS's lack of a ``redshift desert'' further eliminates the need for expensive, space-based\footnote{Ground-based near-IR spectroscopy is too contaminated by
sky-line emission to provide spec-$z$s at the required level of completeness \citep{newman15}.} near-IR spectroscopy to train photo-$z$s with $z > 1.5$.  Highly accurate photo-$z$s will enable science applications that go beyond cosmology.

\begin{wrapfigure}{r}{0.55\textwidth} %
\includegraphics[width=0.55\textwidth]{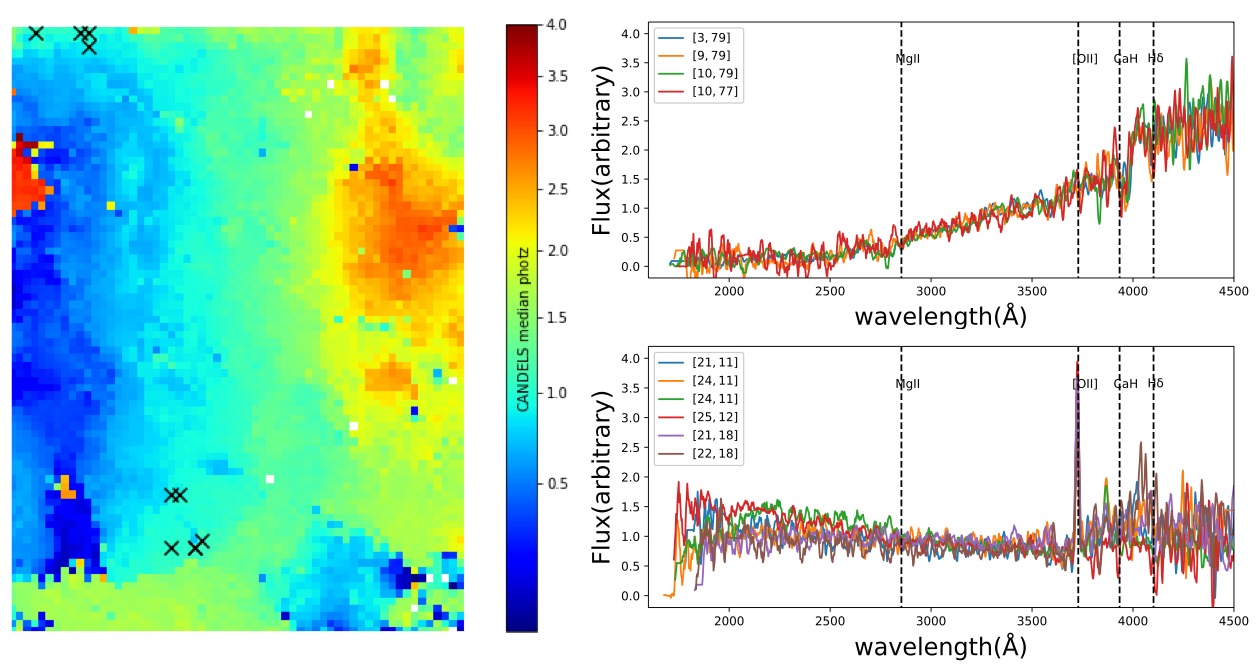}
\caption{\small {\it Left}: A Self-Organizing Map
\citep[SOM;][]{1990Natur.346...24K} from \citet{hemmati18} relating
LSST+WFIRST-like galaxy colors to redshift, $z$. FOBOS spectroscopic
training samples can be optimally designed to populate/calibrate
sparsely sampled regions. {\it Right}: Galaxy spectra from VVDS
\citep{2005A&A...439..845L} in SOM locations marked by black crosses.
More than just redshifts, the detailed similarity of spectral
features of galaxies localized within the SOM demonstrates
higher-level inferences (e.g. SFR) are possible given appropriate
training samples from FOBOS.} \label{fig:SOM} \end{wrapfigure}

\subsubsection{Cosmology with LBG--CMB cross correlation.}
\label{sec:LBG}

High-S/N CMB maps from next-generation CMB observatories (e.g., Simons Observatory and CMB-S4) will provide a cosmic
``reference background'' for measurements of gravitational lensing induced by matter along the line of sight.  After
cross-correlating with Lyman Break Galaxy (LBG) samples, a relatively flat lensing ``kernel'' with power at $z = 2$--5
enables powerful constraints on the Inflation-sensitive matter power spectrum, Horizon-scale General Relativity, cosmic
curvature and neutrino masses, and early Dark Energy \citep{ferraro19}.  \citet{wilson19} explore these constraints in
detail and highlight the need for spectroscopic determination of accurate redshift distributions for the employed LBG
samples. FOBOS would address this need in two ways.  First, several deep-drilling fields targeting $\sim$1000 LBGs BX,
$u$, $g$, and $r$ drop-out candidates per pointing ($\sim$10,000 deg$^{-2}$) would establish the interloper rate and
intrinsic redshift distribution of LBG samples to sufficient precision (this program would likely overlap with the
photo-$z$ program described above).  Second, $\sim$200 LBGs per pointing (2000 deg$^{-2}$) could be included as a
background program when FOBOS observes other sources across the sky, eventually building a 50-100 deg$^2$ data set of
sparse high-$z$ spectroscopy for LBG dN/d$z$ calibration via clustering redshifts \citep[see][]{wilson19}.

%\input{Astro2020-DataScience}
%%%%
% -- Data Science
%%%%

\subsection{FOBOS as an ideal spectroscopic training instrument}
\label{sec:datascience}

In all three science areas above, FOBOS offers significant advances by serving as the leading U.S.~facility in providing  
spectroscopic \emph{training} data. As machine learning
techniques advance in the coming decade, spectroscopic ``training''
as a means of extracting the maximum information from billions of
photometric sources across thousands of deg$^2$ becomes increasingly
important. The required training samples at LSST
depths will fill even the modest Keck focal plane with 10's of
thousands of sources. The key requirements are sensitivity and
multiplex (not field-of-view). Emphasizing these, FOBOS will be an
ideal training facility for LSST-era photometric redshifts
\citep[see][]{salvato19}, galaxy physical properties
\citep[e.g.,][]{davidzon19}, and stellar parameters
\citep[e.g.,][]{2018arXiv180401530T}.

%\input{Astro2020-Description}
%%%%
% -- Instrument Description
%%%%

%%%%%%%%%%%%%%%%%%%%%%%%%%%%%%%%%%%%%%%%%%%%%%%%%%%%%%%%%%%%%%%%%%%%%%%%
\begin{figure}[h!]
%\vskip -0.1in
%\includegraphics[width=\textwidth]{FOBOS_FocalPlane.pdf}
\includegraphics[width=0.8\textwidth]{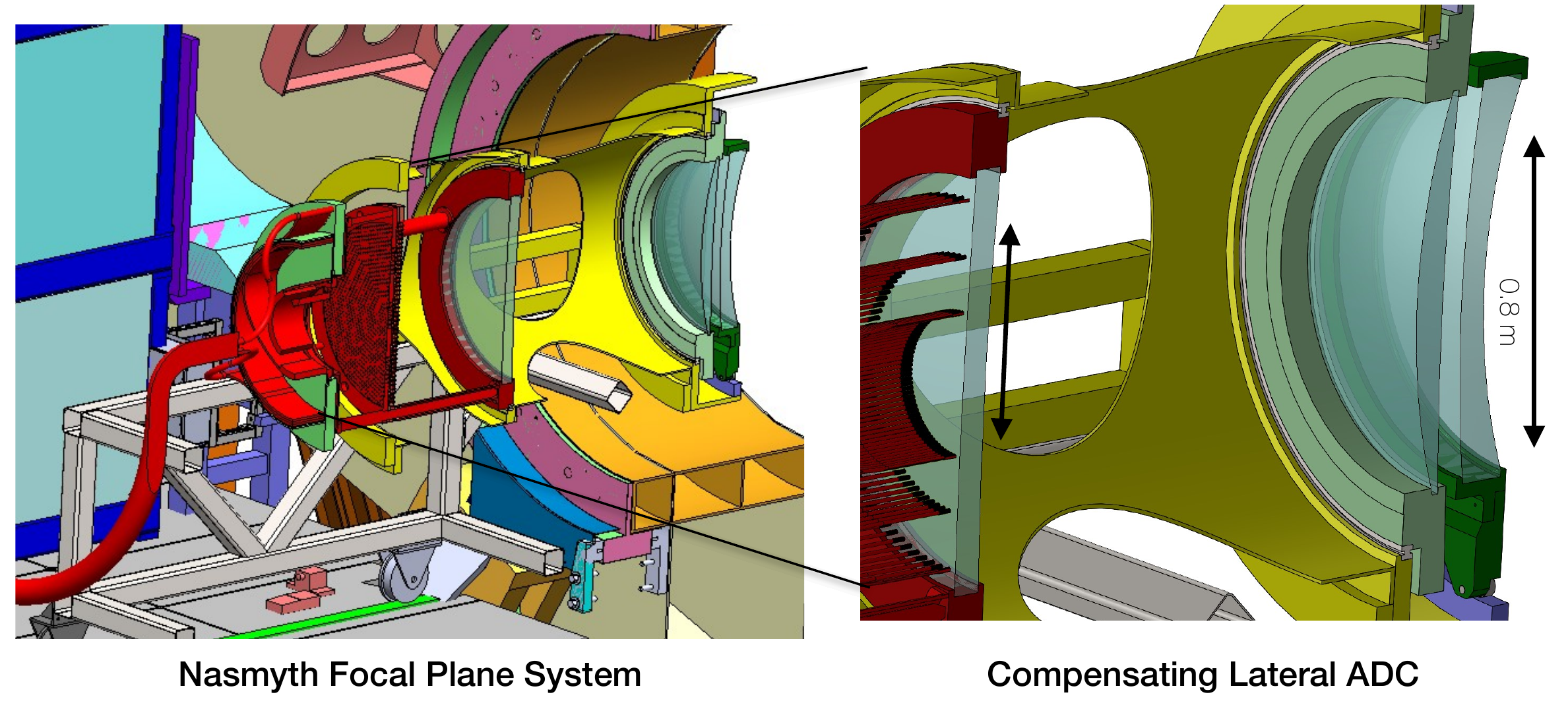}
\caption{\small {\it
Left}: Rendering of FOBOS focal plane system deployed at the Keck II
Nasmyth port. {\it Right}: Rendering of the ADC and focal surface with
Starbugs mounted (red cylinders).}
\label{fig:focalplane}
\end{figure}
%%%%%%%%%%%%%%%%%%%%%%%%%%%%%%%%%%%%%%%%%%%%%%%%%%%%%%%%%%%%%%%%%%%%%%%%

\section{Technical Description}
\label{sec:concept}
% \noindent \comment{1 page}

% Here's an alternative way to put in figures if we want captions on the side (to save space)
% Could introduce a new ``counter'' to count and label figures appropriately
%\centerline{\hbox{\includegraphics[width=0.6\textwidth, angle=0]{FOBOSatKeck_v1.pdf}
%    \hspace{0.1cm} \vspace{2in}
%    \parbox[b]{0.3\textwidth}{\small {\bf Figure ??:} Rendering of FOBOS instrument systems deployed at the Keck II Nasmyth port.  \vspace{2cm}}}}

Mounted at the Nasmyth focus of Keck II Telescope at WMKO, FOBOS\footnote{\url{fobos.ucolick.org}} will
be one of the most powerful spectroscopic facilities deployed in the
next decade. FOBOS includes a compensating lateral atmospheric
dispersion corrector (CLADC; Fig.~\ref{fig:focalplane}) to ensure
that target light from all wavelengths falls on allocated fibers
while also correcting image aberrations at the edges of the 20~arcmin
diameter Keck field. Each of the CLADC lenses is $\sim$700~mm in
diameter, the first two are closely spaced with lateral relative
motions of element one supplied by a single axis of motion acting
along a curve equal to the radius of curvature of the first lens
surface (1028~mm). The total offset of this lens is rather small at
$\sim1$00~mm. The final CLADC lens translates by $\sim$50~mm and
tilts slightly to track the focal-plane shift. This last element acts to
correct the telecentricity error into the fiber system and acts as the
drive surface for the Starbug positioning system.
Starbugs patrol a large on-sky area ($\sim$3~arcmin), enabling
flexible and dynamic targeting configurations with adjacent fibers as
close as 10~arcsec.

Starbugs, first proposed in 2004 \citep{2004SPIE.5495..600M}, and
later perfected by AAO for use on TAIPAN \citep{2016SPIE.9912E..1WS}
are a truly remarkable fiber-positioning system. They move by {\it
walking} on the focal plane using a pair of piezo tube actuators. A
weak vacuum adheres the Starbugs to the surface of the field plate
and provides the frictional normal forces needed to allow for the
walking action of the piezo tubes. Positional feedback is provided by
way of a camera imaging back-illuminated fibers on the focal plane.
This system allows for a highly configurable focal plane both in
terms of target densities and configuration of the fibers within an
individual actuator. The Starbug payload can be both a single fiber or
fiber-bundle IFU. The TAIPAN instrument,
currently on sky conducting a large galaxy survey, is the proving ground
for the readiness of this technology. It is worth noting that,
although Starbugs are our preferred and baseline positioning
technology, no aspect of FOBOS's current front-end design precludes
using a zonal system, such as those used for MOONS, PFS, or DESI. The
last element of the CLADC can be eliminated and replaced with a zonal
actuator bed that conforms to the focal plane shape. Telecentricity
can be maintained by alignment of the actuator axis to the incoming
beam as is currently being designed for the SDSS-V robotic
focal-plane system.

A total of 1800 fibers with 150-$\mu$m core diameter are deployed at the curved focal plane. Microlens fore-optics
convert the f/15 Keck input beam to a faster f/3.2 focal ratio, which both demagnifies the entrance aperture
($\approx$0.9$^{\prime\prime}$ diameter) and allows for a better coupling to the fiber numerical aperture that
minimizes losses from focal ratio degradation (FRD).  With the IFU mode deployed, a different set of lenslet array
fore-optics would more finally sample the focal plane ($\approx$0.33$^{\prime\prime}$.  Accounting for 100 individual
sky fibers and 10 7-fiber flux calibration bundles, 25 science IFUs, each comprised of 61 fibers, would deploy in this
mode.  The diameter of each 61-fiber bundle corresponds to 3$^{\prime\prime}$ on-sky.  A single 61-fiber bundle would remain fixed at the field center in all FOBOS observing modes, enabling rapid target acquisition of transient sources.

The focal-plane plate rotates and translates to follow image
positions as the telescope tracks across the sky. The fiber run is
kept as short as possible to maintain high throughput at UV
wavelengths (a 10~m Polymicro Silica fiber transmits $\sim$70\% and
$\sim$85\% of light at 310~nm and 350~nm, respectively). Special care
is given to stress-relief cabling to minimize instabilities (e.g.,
variable FRD) over the fiber run. To maintain the highest possible
transition efficiency there are no connectors used within the fiber
run. When FOBOS is not in use, the focal-plane unit detaches from the
front-end, ADC module and its associated robotics, and it is stored
with the spectrographs on the Nasmyth platform. This allows the ADC
module to be transferred to any of the instrument park positions. All
other Keck-II instruments can still be used without modification.

%%%%%%%%%%%%%%%%%%%%%%%%%%%%%%%%%%%%%%%%%%%%%%%%%%%%%%%%%%%%%%%%%%%%%%%%
\begin{figure}[h!]
\vskip -0.1in
\includegraphics[width=0.96\textwidth]{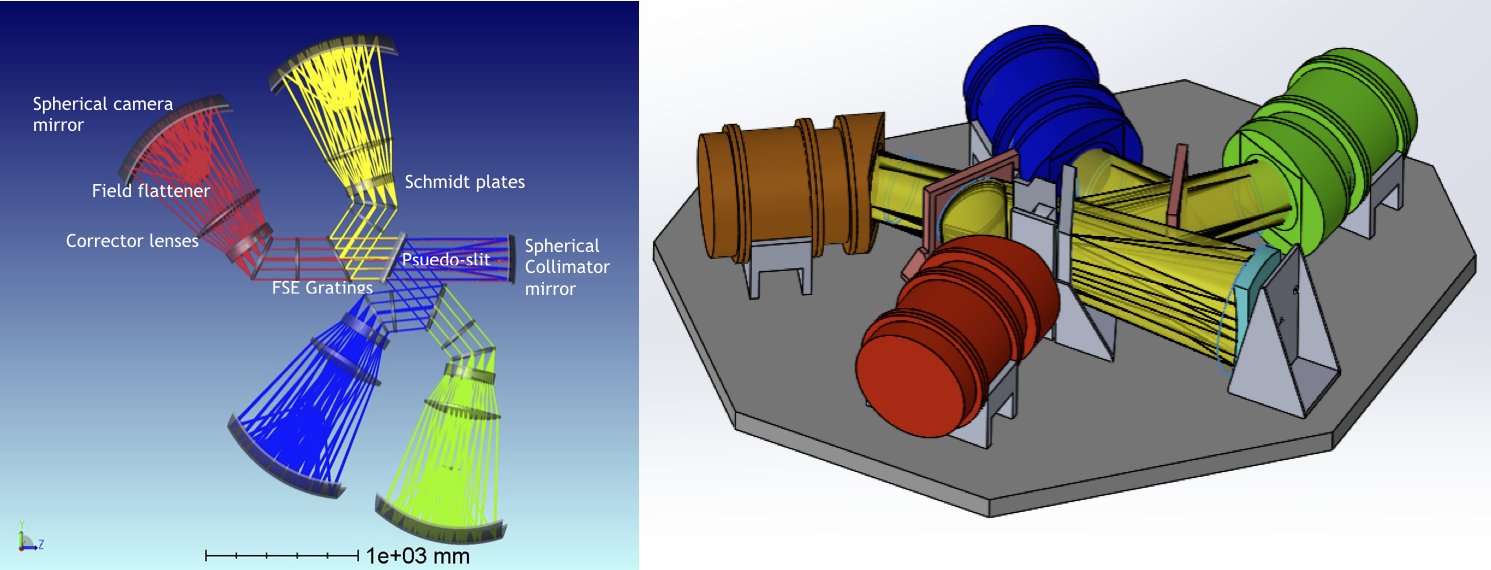}
\caption{\small Optical design (left) and mechancial rendering (right) of a 4-channel FOBOS spectrograph employing catadioptric cameras.
Light from a 600-fiber pseudo-slit strikes a collimating mirror and then
passes back through subsequent dichroics before entering each grating-camera unit.}
\label{fig:spectrograph}
\end{figure}
%%%%%%%%%%%%%%%%%%%%%%%%%%%%%%%%%%%%%%%%%%%%%%%%%%%%%%%%%%%%%%%%%%%%%%%%

FOBOS's three identical spectrographs (Fig.~\ref{fig:spectrograph})
are each fed by a pseudoslit of 600 fibers. Each FOBOS spectrograph
uses a series of dichroics to divide the 259~mm collimated beam into
four wavelength channels, providing an instantaneous broad-band
coverage from 0.31--1 $\mu$m. Fused-silica etched (FSE) gratings
provide mid-channel spectral resolutions of $R\sim3500$ at high
diffraction efficiency in each channel. The dispersed light is
focused by an f/1.1 catadioptric camera\footnote{Based on the camera
design for the Multi-Object Optical and Near-infrared Spectrograph
(MOONS) on the Very Large Telescope (VLT).} and recorded by an
on-axis 4k$\times$4k CCD mounted at the center of the first camera
lens element. Unlike the mountable ADC module, the spectrographs are
housed in a permanent temperature-controlled structure on the Nasmyth
deck. The end-to-end instrument throughput peaks at 60\% and is
greater than 30\% at {\it all} wavelengths.

FOBOS will include observatory level systems for precise instrument
calibration using dome-interior screen illumination, a metrology system
for accurate fiber positioning, and guide cameras for field acquisition
and guiding.  Initial deployment of the focal-plane will focus on a
single-fiber format, with a secondary deployment of multi-format fiber
bundles.  Beyond FOBOS, future instruments could share the focal plane, integrating their
fiber feeds to separate spectrographs optimized for higher spectral resolution, and/or
different wavelengths (e.g., the near-IR).  FOBOS is ideal for taking full advantage of a future
Ground-Layer Adaptive Optics capability.  It also allows for testing advances in spectrometer design that may be critical to future spectroscopic facilities.

% \input{Astro2020-Programmatic}
%%%%
% -- Proposed Work and Budget
%%%%

\subsection{Technology Drivers}
\label{sec:design}

FOBOS will provide key capabilities in the near-term thanks to
deployment at the existing Keck II telescope. It both carries a
relatively modest cost compared to other proposed large-scale
spectroscopic facilities (e.g., MSE, SpecTel) and helps lay the
groundwork for their realization. Thus, while FOBOS will prove to be
a valuable long-term investment for the W.~M.~Keck Observatory, it
can also provide for invaluable technological development leading to
efficiency and cost-cutting strategies for these larger facilities.

\subsubsection{Starbugs fiber positioners} Starbugs are a positioning
technology developed and deployed by Australian Astronomical Optics (AAO), which has partnered with our team to
generate a conceptual design for use of Starbugs by FOBOS. The Starbugs positioning systems is highly attractive
because of its flexibility both in terms of configuring a given set of fibers, as well as the prospect of exchanging
different groups of Starbugs with different payloads and/or those that feed different spectrographs (e.g., high vs.\
low resolution). With such a flexible focal plane deployment, FOBOS can serve as a platform for cost-cutting technology
development, which is not possible with fixed-format instruments like PFS and DESI. Starbugs are currently being tested
on-sky with the TAIPAN instrument at the UK Schmidt Telescope and published results on their performance are expected
in summer 2019.

\subsubsection{Data Systems} A key to FOBOS's success with the
development of robust data-reduction and data-analysis pipelines,
building on the heritage of efforts within SDSS, DESI, and MaNGA. In
particular, the FOBOS data-analysis pipeline (DAP) will take
advantage of the fixed spectral format and common target classes to
provide high-level data products, including Doppler shifts,
emission-line strengths, and template continuum fits (cf., Westfall
et al.; SDSS-IV MaNGA DAP). Planning will include development of
user-friendly platforms built on the Keck Observatory Archive for
serving raw data, reduced spectra, and DAP science products.

\subsection{Current Status} FOBOS is currently in its conceptual design
phase, building from a down-selection process as one of the designs for
the Wide-Field Optical Spectrograph for TMT. Recently, FOBOS has been
awarded Phase-A funding from WMKO Observatory, receiving a full
design-phase endorsement from the Keck Science Steering Committee. These
funds are devoted to completing the conceptual design in preparation for
future funding proposals, particularly the NSF MSIP and MsRI calls.

\subsection{Cost Estimates and Schedule}

Cost estimates for FOBOS reflect its current development phase and are reported in Table \ref{tab:cost} where
available.  Our conceptual-design-phase estimates have higher fidelity for the near-term phases of Preliminary Design
and the beginning of Final Design.  Where possible, costing efforts are based on quotes, and labor efforts are based on
experience with similar systems developed by the institution responsible for a given sub-system.  Our cost projections
in combination with experience from other instruments place FOBOS in the category of a medium-scale ground-based
instrument program suitable for MsRI-2 construction funding. 

Nearly all costs prior to Final Design 2 are allocated to design
efforts; however, a small amount is devoted to prototyping needed to
mitigate risk in key systems.  Our project execution plan divides Final
Design into two phases to allow for a gate for long-lead-time contracts
after a review in June of 2024, such as procurement and construction of
the ADC optics.  The Integration phase also overlaps with the
construction phase to allow for the facility system build-out at Keck
Observatory and to allow for a phased deployment of the multiplexed
focal-plane system.  A full project review is scheduled at the end of
each design phase.  A pre-ship review will be held during integration
prior to delivery of the first spectrograph.   Smaller sub-system
reviews will be held as required.

\begin{table}[h!]
\centering
\footnotesize
\caption{Nominal Schedule and Cost Estimates for FOBOS}
\label{tab:cost}
\vspace*{-10pt}
\begin{tabular}{l | l l r l }
\hline
Phase              &  Start  &     End &         Cost & Fidelity \\
                   &         &         &  (2019 USD) &  \\
\hline
\hline
Conceptual Design  & Q2 2018 & Q1 2021 &   \$730k & Resource-loaded schedule with progress tracking \\
Preliminary Design & Q2 2021 & Q2 2023 &  \$2.6M & Resource-loaded schedule \\
Final Design One   & Q3 2023 & Q2 2024 &  \$1.5M & High-level tasks with cost/effort estimates \\
Final Design Two   & Q3 2024 & Q2 2025 &  \$1.5M & High-level tasks with cost/effort estimates \\
Construction       & Q3 2025 & Q1 2027 & TBD & Block schedule with low-fidelity cost/effort estimates \\
Integration        & Q2 2026 & Q3 2027 & TBD & Block Schedule with low-fidelity cost/effort estimates \\
Commissioning      & Q4 2027 & Q1 2028 &   TBD & Block Schedule with low-fidelity cost/effort estimates \\
\hline
\end{tabular}
\end{table}

% \pagebreak
%\clearpage
\begin{multicols}{2}
\scriptsize
\bibliographystyle{apj}
\bibliography{references}
\end{multicols}

% \textbf{References}

\end{document}